\documentstyle[12pt,aasms]{article}

\newcommand{\EQA}{\begin{eqnarray}}
\newcommand{\ENA}{\end{eqnarray}}
\newcommand{\EQ}{\begin{equation}}
\newcommand{\EN}{\end{equation}}
\newcommand{\fc}{\frac}

\newcommand{\Gam}[2]{\Gamma^{#1}_{{#2}{\theta}}}

\def\bref{\par\noindent\hangindent 20pt}

\def\MN{{ MNRAS \/}}
\def\AnA{{ A\&A \/}}

\newbox\grsign \setbox\grsign=\hbox{$>$}
\newdimen\grdimen \grdimen=\ht\grsign
\newbox\laxbox \newbox\gaxbox
\setbox\gaxbox=\hbox{\raise.5ex\hbox{$>$}\llap
     {\lower.5ex\hbox{$\sim$}}}\ht1=\grdimen\dp1=0pt
\setbox\laxbox=\hbox{\raise.5ex\hbox{$<$}\llap
     {\lower.5ex\hbox{$\sim$}}}\ht2=\grdimen\dp2=0pt

\def\etal{{et al.\ }}

\begin{document}

\title{ACCRETION DISKS AROUND KERR BLACK HOLES: VERTICAL EQUILIBRIUM 
REVISITED}

\author{M. A. Abramowicz\altaffilmark{1,2,3}, A. Lanza\altaffilmark{3},}

\author{and M. J. Percival\altaffilmark{1,4}}

\altaffiltext{1}{Department of Astronomy \& Astrophysics, G{\"o}teborg
University
and Chalmers University of Technology, S-412~96 G{\"o}teborg, Sweden}

\altaffiltext{2} {Nordita, Blegdamsvej 17, DK-2100, Copenhagen {\O}, Denmark}

\altaffiltext{3}{Scuola Internazionale Superiore di Studi Avanzati,
via Beirut 2-4, I-34014 Trieste, Italy}

\altaffiltext{4}{Mathematical Tripos, Part III, Cambridge University,  
England}

\vskip 5truecm
\noindent 
\leftline {M.A. Abramowicz: marek@tfa.fy.chalmers.se}
\leftline {A. Lanza: lanza@sissa.it}
\leftline {M.J. Percival percival@tfa.fy.chalmers.se}
 
\begin{abstract}
We re-derive the equation for vertical hydrodynamical equilibrium of
stationary thin and slim accretion disks in the Kerr spacetime. All the
previous derivations have been unsatisfactory, yielding unphysical
singularities. Our equation is non-singular, more general, and simpler
than those previously derived. 
\end{abstract}


\section{PHYSICAL MEANING OF THE VERTICAL EQUATION}

\noindent In this paper, we derive, in Kerr geometry, the {\it vertical
equation} for stationary thin and slim accretion disks. The equation
governs the balance between the vertical pressure gradient, vertical
acceleration, and vertical components of all the inertial forces. 
Vertical direction means here the direction orthogonal to the plane of the
disk, for example, the $z$ direction in cylindrical coordinates, or the
$\theta$ direction in spherical coordinates. We assume that the disks are
stationary and axially symmetric and that the vertical thickness of the
disk $H$ is much smaller than the corresponding cylindrical radius $r$. 

\noindent We derive the equation first in Newtonian theory in order to
stress some points that are often overlooked, but the new result presented
in this paper is the fully self-consistent derivation of the vertical
equation in general relativity, in particular for the Kerr spacetime. 

\noindent In the reference frame comoving with the matter, the vertical
pressure gradient force balances the combined vertical components of all
the inertial forces: gravity, centrifugal force, the Euler force due to
the vertical acceleration, and (in the case of a rotating black hole) the
general relativistic Lense-Thirring force (dragging of inertial frames).
The Coriolis force always vanishes in the comoving frame of reference and
does not need to be considered. 

\noindent Previously, the vertical equation has been derived by Novikov \&
Thorne (1973; NT), Riffert \& Herold (1995; RH) and Abramowicz, Chen,
Granath \& Lasota (1996; ACGL). None of these derivations was
satisfactory. They all made incorrect assumptions that introduced
unphysical singularity into the final form of the vertical equation.
Formulae derived by NT and RH are singular at the location of the circular
photon orbit, while the formula derived by ACGL is singular at the
horizon. NT based their derivation on a discussion of the gravitational
tidal forces in the vertical {\it cylindrical} direction. This set a
standard for several further derivations, but (as pointed out by Lasota \&
Abramowicz, 1996) was not a fortunate choice, because a mathematically
correct discussion of the vertical tidal forces at the horizon (not yet
given) would be very complicated and thus impractical. On the other hand,
a direct approach based on the relativistic Euler equation is
straightforward and simple.  Here, we derive the equation directly from
the relativistic Euler equation and make no additional simplifying
assumptions. Our derivation is based on a general discussion given by
Lasota \& Abramowicz (1996).  

\section {THE METHOD: SLIM ACCRETION DISKS}

\noindent Physical quantities $X(r,z)=X(r,-z)$ that are symmetric with
respect to the plane of the disk, $z = 0$, are expanded around the plane
according to $$X(r,z) = \sum_{i=0}^{N=\infty} X_{2i}(r)\left({z\over
r}\right)^{2i}.$$ Physical quantities antisymmetric with respect to the
plane $X(r,z) = -X(r,-z)$ have expansion $$X(r,z) = \left({z\over
r}\right)\sum_{i=0}^{N=\infty} X_{2i}(r) \left({z\over r}\right)^{2i}.$$
Putting these expansions into the partial differential equations that
describe stationary accretion disks, one converts the partial differential
equations into set of $N=\infty$ equations with $N=\infty$ unknowns.
Truncating the expansions at $N=0$, thus keeping only zeroth order terms
in $(z/r)$, would {\it not} give a closed set: in this case there are more
unknown functions than the equations. Truncating at $N=1$ gives a closed
set, with the number of unknown functions equal to the number of
equations.  The same is true for any $N>1$.  Only terms of the order
$(z/r)^0$ and $(z/r)^2$ appear in the $N=1$ equations. Therefore the
approximation based on truncating all expansions at $N=1$ offers two
advanteges: (1) it is mathematically well defined and consistent, (2) it
involves the smallest number of equations to be solved.  The precise
mathematical meaning of the ``slim accretion disks'' approach (Abramowicz
\etal, 1988) is just that: one should keep {\it all} the terms of the
order $(z/r)^2$ and $(z/r)^0$ in {\it all} the equations. This should be
contrasted with the ``thin accretion disks'' approach, where terms of the
order of $(z/r)^2$ are kept in {\it some} of the equations, but in others
they are rejected according to the tradition established in the
influential paper by Shakura \& Sunyaev (1973). It is now well understood
that this traditional approach often causes serious physical
inconsistencies which make the thin disk models inadequate for describing
several astrophysically interesting types of accretion disks. 

\noindent Here, we derive one particular equation~--- the vertical
equation~--- which belongs to the set of the slim accretion disk
equations. All the other equations have been derived in the Kerr geometry
by Lasota (1994) and Abramowicz \etal (1996). The vertical equation is
itself of the order $(z/r)^2$. We keep all the terms of zeroth and second
orders in our derivation. Thus, our final ``slim disk'' equation contains
all the $(z/r)^2$ terms that follow from a mathematically consistent
procedure. However, we also discuss here the ``thin disk'' form of this
equation in which some of the $(z/r)^2$ terms have been rejected in a way
which is not mathematically consistent, but follows the traditional thin
disk approach. 

\section{NEWTONIAN DERIVATION}

\noindent It will be helpful to derive the vertical equation in Newtonian
theory in both cylindrical and spherical coordinates. Of course, the final
result does not depend on the coordinates used. However, there are some
interesting differences between the equations written in cylindrical and
spherical coordinates. When we use cylindrical coordinates, there is no
centrifugal force in the ``vertical'' direction, while when we use the
spherical coordinates, {\it there is no gravitational force in the
``vertical'' direction}. As we will see, it is exactly this property that
makes the spherical coordinates much better adapted for descring the flow
near the black hole horizon. All previous derivations have been carried
out in cylindrical coordinates, and we feel that it would be proper if we
explain our derivation in these coordinates first. 

\vskip 0.3truecm
\centerline{\it 3.1. Cylindrical coordinates}
\vskip 0.3truecm

\noindent Let $H = z(r)$ describe the location of the disk surface in
cylindrical coordinates $(r,z,\phi)$. According to our basic 
assumption that terms up to the second order should be kept, the 
pressure $P(r,z)$, in terms of $({z/r})$, has the expansion,
\begin{equation}
P(r,z) = P_0(r) - \left({z\over r}\right)^2 P_2 (r) + {\cal
O}^4\left({z\over r}\right),
\end{equation}
where the small dimensionless parameter $(z/r)$ gives the distance from
the plane of the disk. The minus sign before $({z/ r})^2P_2(r)$ has been
chosen to make $P_2(r) > 0$.  $P_0(r)$ is the pressure at the plane of the
disk. At the disk surface, $P(r,z)=0$, and therefore
\begin{equation}
P_0(r) = \left({H \over r}\right)^2 P_2(r). 
\end{equation}

\noindent
Note that for $z \sim H$ higher order terms in expansion (1) are not 
numerically small compared with the lower order ones. This is a weak 
point of the method but we have checked that the numerical inaccuracies 
introduced by this are of the same order as those of other approximate 
methods.

\noindent From equations (1) and (2) it follows that,
\begin{equation}
P(r,z) = P_0(r)\left[1 - \left({z\over H} \right)^2\right],~~{\rm and}~~
{{\partial P}\over {\partial z}} = -{{2z}\over H^2}P_0(r).
\end{equation}
We shall denote by $\rho_0(r)$ the density at the plane of the disk. We
{\it do not} assume anything about the vertical distribution of the
density of matter $\rho(r,z)$ as all derivations based on the ``vertical
integration'' do. In particular, we {\it do not} assume that the density
vanishes at the disk surface, or that it is in accordance with a
polytropic equation of state, $P=K\rho^{1+1/n}$, as assumed by Hoshi
(1977). Hoshi's formula reduces to our equation (3) for $n=0$. 

\noindent We shall write for the radial and cylindrical components of
velocity,
\begin{equation}
v_r = v_0(r) + {\cal O}^2\left({z\over r}\right),
\end{equation}
\begin{equation}
v_z = \left({z\over r}\right)u_1(r) + {\cal O}^3\left({z\over r}\right).
\end{equation}

\noindent For stationary flows, the vertical equation has the general
form,
\begin{equation}
{1\over \rho} \left( {{\partial P}\over{\partial z}} \right) + \left(
{{\partial \Phi}\over{\partial z}} \right) + v_r{{\partial
v_z}\over{\partial r}} + v_z{{\partial v_z}\over {\partial z}} = 0.
\end{equation}
Here $\Phi = \Phi(R)$ is the gravitational potential that depends only on
the spherical radius $R = \sqrt{r^2 + z^2},~(\partial R/\partial z) =
z/R$. The Keplerian orbital velocity $V_K(R)$ in the potential $\Phi(R)$
is given by,
\begin{equation}
V_K^2(R) = \left(R{{d\Phi}\over{dR}}\right).
\end{equation}
Taking these relations into account, we write equation (7) in the form,
\begin{equation}
\left({z\over r}\right) \left\{ -2 {P_0\over {\rho_0}} + \left({H\over
r}\right)^2 \left[ V_K^2 + rv_0{du_1\over dr} + u_1^2 - u_1v_0\right]
\right\} + {\cal O}^3\left({z\over r}\right) = 0.
\end{equation}
Thus, by demanding that the expression in the curly brackets vanishes, we
find the most general version of the vertical equation,
\begin{equation}
-2 {P_0\over {\rho_0}} + \left({H\over r}\right)^2 \left[ V_K^2 +
rv_0{du_1\over dr} + u_1^2 - u_1v_0\right] = 0.
\end{equation}
In this most general version, an additional unknown function $u_1(r)$
appears, that is not present in the standard thin or slim set of
equations. The additional equation which closes the set (not needed in
the standard treatment where $u_1=0$ is assumed {\it ad hoc}) is provided
by the surface boundary condition which says that there should be no
component of the total velocity orthogonal to the surface of the disk. In
cylindrical coordinates this means that
\begin{equation}
{{u_1}\over {v_0}} = {{d\ln H}\over {d\ln r}}.
\end{equation}
>From the above equation it follows that $u_1$ and $v_0$ must be of the
same order with respect to $(H/r)$ and that $u_1 \equiv 0$, is only a good
approximation if $H$ varies very slowly with $r$. 

\noindent Therefore, in the most general situation, the vertical
equation consists of {\it two} equations: equation (9) which governs the
vertical balance of forces and acceleration,
and equation (10) which demands that in a
stationary disk there is no component of velocity orthogonal to the
surface.

\noindent From equation (9) it follows that in general $P_0/\rho_0 \sim
C_S^2 \sim (H/r)^2$ with $C_S$ being the sound speed. In addition, it is
obviously $V_K^2 \sim (H/r)^0$. One can make simplifications in equation
(9) by considering the $(H/r)$ order of each term.  In the standard thin
disk model the radial flow is very sub-sonic, and therefore $v_0 \sim u_1
\ll C_S$. Therefore, the terms $rv_0(du_1/dr)$, $u_1^2$, and $u_1v_0$ are
of order higher than $(H/r)^2$ and may be dropped. The standard thin disk
version of the vertical equation therefore has the form,
\begin{equation}
{P_0\over {\rho_0}} = {1\over 2}\left({H\over r}\right)^2 V_K^2.
\end{equation}
In the transonic part of the slim disks the terms containing $u_1$ are
exactly of the $(H/r)^2$ order and may also be neglected. From
consideration of the vertical equilibrium alone it seems possible that the
thin disk version of vertical equation could be also written in the form
(11), without the terms containing $u_1$.  However, if these terms have
been dropped, then in doing calculations with the full set of equations,
one should perform a {\it consistency check}, and calculate the ratios
$(u_1v_0)/C_S^2$ and $(u_1u_1)/C_S^2$ using eqn. (10).  If these ratios are
$\sim 1$, one {\it must} use the general form of the vertical equation,
provided by equations (10) and (9).  This is the case close to the horizon
of the black hole where $v_0 \sim u_1 \sim c \gg C_S$. We shall return to
this point later. 

\vskip 0.3truecm
\centerline{\it 3.2. Spherical coordinates}
\vskip 0.3truecm

\noindent In spherical coordinates ($R,~\theta,~\phi$) the small parameter
that characterizes the distance from the plane of the disk should now be
$\cos \theta$, and the equation for the surface of the disk is given by
$\theta = \Theta (R)$. Obviously, in this case $H/r \approx \cos \Theta$. 
According to this, a formula for the pressure, similar to equation (3)
should be written as,
\begin{equation}
P(R, \theta) = P_0(R)\left[1  - \left( {{\cos^2 \theta}\over  {\cos^2 
\Theta}} \right) \right],~~ 
{\rm and}~~{1\over R}{{\partial P}\over {\partial \theta}} = \fc{2 R \cos
\theta}{H^2}P_0(R).
\end{equation}
For the radial, azimuthal and polar components of the velocity we again
write,
\begin{equation}
v_R(R,\theta) = v_0(R) + {\cal O}^2(\cos \theta),
\end{equation}
\begin{equation}
v_{\phi}(R,\theta) = v_{\phi}(R) + {\cal O}^2(\cos \theta),~~~
v_{\theta}(R,\theta) = \cos \theta u_1(R) + {\cal O}^3(\cos \theta).
\end{equation}
Note, that $v_{\phi}$ {\it does not} have the physical dimension of {\it
velocity}. This is because, in spherical coordinates, $v^{\phi} \equiv
(d\phi/dt) =\Omega$, with $\Omega$ being the angular velocity, and
$v_{\phi} \equiv g_{\phi\phi}v^{\phi} = R^2\sin^2\theta \Omega = r^2\Omega
= {\cal L}$, with ${\cal L}$ being the angular momentum per unit mass.
Similarly, $v^{\theta}$ has the physical dimensions of angular velocity,
and both $v_{\theta}$ and $u_1$ have the dimensions of angular momentum per
unit mass. Of course, $v^R$, $v_R$ and $v_0$ have the dimensions of
velocity. One should bear this in mind when checking the physical
dimensions of our formulae in spherical coordinates, or when comparing
our formulae with those given {\it e.g.} by Tassoul (1978).  Tassoul's
azimuthal velocity ``$v_{\phi}$'', which we denote here (to avoid
confusion) by $V_{(\phi)}$, is given by $V_{(\phi)} = R\sin\theta v^{\phi}
= v_{\phi}/R\sin\theta$, and has the physical dimensions of velocity. 

\noindent In spherical coordinates, the condition that there should be no
component of velocity orthogonal to the surface of the disk, similar to
that given by equation (10), takes the form,
\begin{equation}
{1\over R}{u_1 \over v_0} = {{d\ln \vert {\pi \over2} - \Theta\vert}\over
{d\ln R}}.
\end{equation}

\noindent The general form of the vertical equilibrium is now given by a
formula similar to (7), but written in spherical coordinates,
\begin{equation}
\fc{1}{\rho} \fc{\partial
P}{\partial \theta} +
R v_R \fc{\partial \left(v_\theta/R\right)}{R} +\fc {v_\theta}{R}
\fc{\partial \left(v_\theta/R\right)}{\partial \theta} + \fc{v_R
v_\theta}{R}+\fc {v^2_\phi}{R^2} \cot \theta= 0.
\end{equation}
\noindent From these formulae, using the same procedure as before, we
derive the general form of vertical equation in spherical coordinates,
similar to its cylindrical version (9)
\begin{equation}
-2\fc{P_0}{\rho_0}+\left( \fc{H}{R} \right)^2 \left[{V_{(\phi)}}^2 +
\fc{u_1^2}{R^2}-v_0 \fc{du_1}{dR} \right]=0
\end{equation}
Note a significant difference between the two versions of the same vertical
equation, written in cylindrical coordinates (9), and in spherical
coordinates (17). While in (9) the Keplerian velocity $V_K^2$ appears, in
(17) we have the rotational velocity $V_{(\phi)}^2$. The difference is due
to the fact that in the $z$ direction there is gravity force ($\sim
V_K^2/R$) but no centrifugal force, while in the $\theta$ direction there
is centrifugal force ($\sim V_{(\phi)}^2/R$) but no gravity. {\it Does
this matter?} One may argue that it does not, because
\begin{equation}
V_K^2 - V_{(\phi)}^2 \sim {H^2\over r^2} = \cos^2 \Theta,
\end{equation}
and therefore, with accuracy to the higher order terms, these two versions
of the vertical equation are identical (for the standard thin disk $V_K^2
\equiv V_{(\phi)}^2$). 

\noindent But actually it {\it does matter a lot}. Firstly, in both
Newtonian and general relativistic theories, the stationary accretion
flows calculated in realistic 2-D and 3-D simulations are much more
resembling quasi spherical flows ($\Theta \approx {\rm constant}$) than
quasi horizontal flows ($H \approx {\rm constant}$). This has been noticed
by several authors, {\it e.g.} by Papaloizou \& Szuszkiewicz (1994), and
Narayan \& Yi (1995). Thus, while $u_1 = 0$ may be a reasonable
approximation in spherical coordinates, it is not so in the cylindrical
ones. Secondly, anticipating the relativistic discussion in the next
section, the Keplerian velocity $V_K(R)$ is singular at the location of
the circular photon orbit, while the azimuthal velocity $v_{\phi}$ is
non-singular everywhere, including the horizon. Thirdly, we know {\it
exactly} what happens on the horizon in spherical coordinates. A general
theorem (Carter, 1973) ensures that in spherical coordinates,
$u_1(R_H)=0$.  The use of cylindrical coordinates to describe the flow
brings artificial, and technically complicated, difficulties near the
horizon. These are generated only by a bad choice of coordinates (as, 
e.g. in Riffert \& Herold, 1995). 

\section{RELATIVISTIC DERIVATION}

\noindent We shall now derive the vertical equation in Kerr geometry, 
using the same method as in the previous Section. It is convenient  
to work at the beginning with a general metric,
\begin{equation}
ds^2 = g_{tt}dt^2 + 2g_{t\phi}dtd\phi + g_{\phi \phi}d{\phi}^2 + g_{RR}dR^2 
+ g_{\theta \theta}d{\theta}^2,
\end{equation}
and specify the Kerr metric functions $g_{ik} = g_{ik}(R,\theta)$ at the end 
of the calculations.

\noindent The stress-energy tensor for a perfect fluid has the form,
\begin{equation}
T^{i}_{~k}= \rho u^i u_k-\delta^i_{~k} P,
\end{equation}
where $u^i = dx^i/ds$ is the four-velocity of the fluid, and $\rho =
\varepsilon + P$, where $\varepsilon$ is the energy density (we use units
in which $c = 1 = G$). The four-velocity is a unit vector, which means
that
\begin{equation}
1 = g^{ik}u_iu_k = g^{tt}u_tu_t + 2g^{t\phi}u_tu_{\phi} + g^{\phi
\phi}u_{\phi}u_{\phi} + g^{RR}u_Ru_R + g^{\theta
\theta}u_{\theta}u_{\theta}.
\end{equation}
The inverse metric $g^{ik}$ obeys $g^{ik}g_{jk} = \delta^i_{~j}$. 

\noindent We derive the vertical equation taking the $j=\theta$ component
of the general equation $h^i_{~j}\nabla_k T^k_{~i} = 0$. Here $\nabla_k$
is the covariant derivative operator, and $h^i_{~j} = \delta^i_{~j} - u^i
u_j$ is the projection tensor. This gives, at the end of simple standard
calculations,
\begin{equation}
{1\over {\rho}}{{\partial P}\over {\partial \theta}} = u^k{{\partial
u_{\theta}}\over {\partial x^k}} - \Gamma_{~\theta k}^i u_i u^k,
\end{equation}
where $\Gamma_{~\theta k}^i$ is the Christoffell symbol that should be
computed from the metric components and their first derivatives.  This
relativistic equation provides the replacement for the Newtonian equation
(16). As in the Newtonian case, the leading order of this equation is
$(\cos \theta)^1$. Thus, in calculations that follow, we shall keep only
these terms, and neglect the ${\cal O}^3(\cos \theta)$ and higher order
terms. 

\noindent The $-\Gamma_{~\theta k}^i u_i u^k$ term should be computed with  
the help of equation (21). The result in Kerr geometry is
\begin{eqnarray}
-\Gamma_{~\theta k}^i u_i u^k =- u_{\phi}u_{\phi}\left[g^{\phi \phi}
\Gam{\phi}{\phi}+g^{t \phi} \Gam{\phi}{t}-g^{\phi \phi} \Gam{R}{R}\right]
\nonumber \\
- u_{\phi}u_{t}\left[g^{tt} \Gam{\phi}{t}+g^{t \phi} 
\Gam{\phi}{\phi}+g^{t \phi} \Gam{t}{t}+g^{\phi \phi}
\Gam{t}{\phi} -2 g^{t \phi} \Gam{R}{R}\right] \nonumber \\ 
-u_{t}u_{t}\left[ g^{tt} \Gam{t}{t}
+g^{t \phi} \Gam{t}{\phi}-g^{tt} \Gam{R}{R}\right]-\Gam{R}{R}.
\end{eqnarray}
In deriving this equation we took into account that $u_\theta 
u_\theta\sim \cos^2 \theta$ should be dropped, and that in Kerr geometry 
$g^{RR}\Gamma^\theta_{\theta R}+g^{\theta\theta}\Gamma^R_{\theta\theta}=0$.
The term $u^k{{\partial u_{\theta}}/{\partial x^k}}$ equals
\begin{equation}
u^k{{\partial u_{\theta}}\over {\partial x^k}}
=\cos \theta \left[g^{RR}V_0 \fc{dU_1}{dR}-g^{\theta \theta}U_1^2\right],
\end{equation}
where $U_1$ and $V_0$ are defined in terms of the following expansions 
[{\it cf.} equations (13) and (14)],
\begin{equation}
u_R = V_0(R) + {\cal O}^2(\cos \theta),~~
u_{\theta} = \cos \theta U_1(R) + {\cal O}^3(\cos \theta).
\end{equation}
One also needs to use the relations $u^{\theta} = g^{\theta 
\theta}u_{\theta}$ and $u^R = g^{RR}u_R$.

\noindent The terms containing pressure gradient should be treated the 
same way as in the Newtonian derivation. 

\noindent To the desired orders, the components of the inverse metric, and
the Christoffel symbols which appear in (23) and (24) are in Kerr
geometry,
\begin{equation} \begin{array}{llll}
g^{tt}&=&\fc{2a^2(R+2M)+R^3}{R\Delta}, &g^{t\phi}=\fc{2aM}{R\Delta}, \\
g^{\phi \phi}&=&-\fc{R-2M}{R \Delta }, &g^{RR}=-\fc{\Delta}{R^2}, \\
g^{\theta\theta}&=&-\fc{1}{R^2}, &\Gam{t}{t}=-\fc{2a^2M}{R^3}\cos\theta,
\\ \Gam{\phi}{t}&=&-\fc{2aM}{R^3}\cos\theta,
&\Gam{t}{\phi}=\fc{2a^3M}{R^3}\cos\theta, \\
\Gam{\phi}{\phi}&=&\left(1+\fc{2a^2M}{R^3}\right)\cos\theta,
&\Gam{R}{R}=-\fc{a^2}{R^2}\cos\theta
\end{array} \end{equation}
By inserting these into equations (22 - 24), we obtain the final result,
\begin{equation}
-2{{P_0}\over {\rho_0}} + \left({H\over
R}\right)^2\left(\fc{1}{R^2}\right) \left[{\cal L}^2_{*}+U_1^2-\Delta V_0
\fc{dU_1}{dR} \right]= 0.
\end{equation}
Here ${\cal L}_{*}^2 = {\cal L}^2-a^2 ({\cal E}-1)$, where ${\cal L} =
-u_{\phi}$ is the conserved angular momentum for a geodesic motion and
that ${\cal E} = u_t$ is the conserved energy for such a motion. Realistic
matter almost free-falls (moves along geodesics lines) when it crosses the 
horizon. Thus, ${\cal L}_* \approx const$ near the horizon. 

\noindent Formula (27) above gives the most general version of the
vertical equation for stationary flow. On the horizon, $U_1=0$, and
${\cal L}_{*}$ is finite, $V_0 \rightarrow \infty$, but $\Delta
\rightarrow 0$, in such a way that $V_0\Delta$ is a finite quantity. Thus
our equation is non-singular on the horizon. 

\noindent As in Newtonian theory, equation (27) should be considered
together with equation (15) which, obviously, has the same form in
relativity. However, in the spirit of the ``thin disk approach'', we
suggest that in practical applications, involving stationary flow, it is
always {\it safe} to drop the terms containing $U_1$. As we have already
explained, they cannot be significant anywhere in the flow: far from the
black hole because the flow there is sub-sonic or transonic, and close to
the hole because flow there is quasi spherical.  Thus, we conclude, that
the vertical equation could  be taken in practical applications in the
form,
\begin{equation}
-2{{P_0}\over {\rho_0}} + \left({H\over R}\right)^2
 {{{\cal L}^2_{*}}\over R^2} = 0.
\end{equation}
The Schwarzschild version of this equation was first derived by Lasota and 
Abramowicz (1996).

\noindent Most of this work was done when MAA and MJP were visiting SISSA
in the summer of 1996. MAA acknowledges the partial support by the
Nordita's Nordic Project on {\it Non-Linear phenomena in accretion disks
around black holes}. The work of AL has been supported by MURST. We thank 
Igor Igumenschev, Jean-Pierre Lasota, John Miller, Luciano Rezzola, and 
the referee for helpful comments.


\begin{references}

\bref Abramowicz M.A., Czerny B., Lasota J.-P., \& Szuszkiewicz E. 1988, \apj 
332, 646
\bref Abramowicz, M. A., Chen, X., Granath, B.  \& Lasota, J.-P., 1996, \apj, 
in press
\bref Carter, B., 1973, in Black Holes, eds. C. de Witt \& B.S. de Witt
(New York: Gordon \& Breach) 58
\bref Hoshi, R., 1977, Prog. Theor. Phys., 58, 1191
\bref Lasota, J.-P. 1994, in Theory of Accretion Disks 2., eds. Dushl 
W.J., Frank J., Meyer F., Meyer-Hofmeister E. \& Tscharnuter W.M. 
(Dordrecht: Kulwer)
\bref Lasota, J.-P., Abramowicz, M.A. 1996, Class. Quantum Grav., in press 
\bref Narayan, R., \& Yi, I. 1995, \apj, 444, 231
\bref Novikov I.D. \& Thorne K.S. (NT), 1973,  in Black Holes, 
ed. C. de Witt \& B.S. de Witt (New York: Gordon \& Breach) 343
\bref Papaloizou, J. \&  Szuszkiewicz, E., \MN, 268, 29
\bref Riffert, H. \& Herold, H., 1995, \apj, 450, 508
\bref Tassoul, J.-L. 1978, Theory of Rotating Stars, Princeton University 
Press, Princeton (pp 479-480) 
\bref Shakura, N.I.,  \& Sunyaev, R.A. 1973, \AnA, 24, 337
\end{references}
\end{document}